\documentclass[prd,reprint,nofootinbib]{revtex4-1}
\usepackage{CJK}

\usepackage{amssymb} 
\usepackage{amsmath}
\usepackage{xfrac}
\usepackage{bm}
\usepackage{rotating}
\usepackage{subcaption}
\usepackage{caption}
\usepackage{graphicx}
\usepackage{datetime}
\usepackage{braket}
\usepackage[noabbrev]{cleveref}

\renewcommand{\L}{\mathcal{L}}
\newcommand{\G}{\mathcal{G}}
\newcommand{\N}{{\cal N}}
\newcommand{\hf}{\frac{1}{2}}
\newcommand{\fb}[2]{\left(\frac{#1}{#2}\right)}
\newcommand{\abs}[1]{\left|#1\right|}

\DeclareMathOperator{\adj}{adj}

\crefrangeformat{equation}{(#3#1#4)-(#5#2#6)}

\newdateformat{monthyeardate}{%
  \monthname[\THEMONTH] \THEYEAR}

\begin{document}

\begin{flushright}
\hspace{1em} \\MAN/HEP/2019/012
\\
\monthyeardate{\today}
\end{flushright}

\title{\Large{Initial Conditions of Inflation in a Bianchi I Universe}}

\author{Kieran Finn}\email{kieran.finn@manchester.ac.uk}

\affiliation{Department of Physics and Astronomy, University of Manchester, Manchester
 M13 9PL, United Kingdom}

\begin{abstract}
We investigate the initial conditions of inflation in a Bianchi~I universe that is homogeneous but not isotropic. We use the Eisenhart lift to describe such a theory geometrically as geodesics on a field space manifold. We construct the phase-space manifold of the theory by considering the tangent bundle of the field space and equipping it with a natural metric. We find that the total volume of this manifold is finite for a wide class of inflationary models. We therefore take the initial conditions to be uniformly distributed over it in accordance with Laplace's principle of indifference. This results in a normalisable, reparametrisation invariant measure on the set of initial conditions of inflation in a Bianchi~I universe. We find that this measure favours an initial state in which the inflaton field is at or near its minimum, with a mild preference for some initial anisotropy. Since inflation requires an initial field value with a large displacement from its minimum, we therefore conclude that the theory of inflation requires finely tuned initial conditions.
\end{abstract}

\maketitle

\section{Introduction}
Inflation~\cite{Guth:1980zm,Linde:1981mu,Albrecht:1982wi} is the leading theory describing the early Universe. In particular, inflation is often invoked as the solution to the classic cosmological puzzles of the horizon and flatness problems. Both of these are problems of fine tuning of the initial state of the Universe. There is nothing preventing a completely flat and homogeneous universe arising in the Hot Big Bang model, but it requires incredibly specific initial conditions. It is therefore imperative that inflation require less finely tuned initial conditions if it is to solve these problems satisfactorily.

There has been much debate in the literature as to whether the initial conditions required for inflation are finely tuned~\cite{Linde:1985ub,Chmielowski:1988zb,Linde:1993nz,Linde:1993xx,GarciaBellido:1993wn,Coule:1994gd,Cho:1994vy,Stoeger:2004sn,Garriga:2005av,Page:2006er,Aguirre:2006na,Aguirre:2006ak,Clifton:2007bn,Winitzki:2008yb,Ijjas:2013vea,Guth:2013sya,Corichi:2013kua,Kleban:2016sqm,Clough:2016ymm,Clough:2017efm,Aurrekoetxea:2019fhr}. However, there is still no consensus with some authors claiming that inflation happens generically~\cite{Gibbons:1986xk,Goldwirth:1991rj,East:2015ggf,Linde:2017pwt}, while others argue precisely the opposite~\cite{Penrose:1988mg,Gibbons:2006pa, Ijjas:2014nta, Berezhiani:2015ola}.

The main reason for the differences in conclusion is the infinite size of the space of allowed initial conditions. This infinity must be regulated in order to obtain a finite number for the likelihood of inflation and, as shown by~\cite{Hawking:1987bi,Sloan:2015bha}, the result one obtains can depend strongly on the choice of regulator.

In~\cite{Finn:2018krt} (hereafter F18) we constructed a measure of initial conditions that is finite for a large class of inflationary models and thus has no need for regularization. We achieved this by using the Eisenhart lift~\cite{Eisenhart,Finn:2018cfs} to describe inflation as the geodesic motion on a field-space manifold. We then took as our measure of initial conditions the diffeomorphism invariant volume element of the tangent bundle of that manifold. We found the total volume of the tangent bundle to be finite provided the inflationary potential diverges faster than $\varphi^2$ as $\varphi\to\infty$ and there exists a non-zero cosmological constant. We were thus able to normalise our measure without the use of a regulator.

In this Letter we extend the applicability of~F18, whose results were calculated for a flat, homogeneous and isotropic universe. We still assume the Universe to be flat and homogeneous, but shall relax the assumption of isotropy and thus consider a Bianchi~I universe~\cite{Bianchi}. Although we do not observe anisotropy in the Universe today, there is no reason to assume it was not present in the early Universe. Indeed, smoothing out initial anisotropy is one of the key achievements of inflation~\cite{Collins:1972tf,Moss:1986ud,Barrow:1986jd}. The aim of this Letter is therefore to see whether allowing for such initial anisotropy changes the results of~F18.

\section{The Eisenhart Lift}
We begin by briefly reviewing the Eisenhart lift~\cite{Eisenhart,Finn:2018cfs} and showing how it can be used to describe scalar field theories geometrically. Consider a theory with $N$ degrees of freedom, labelled by $\varphi^i$ (collectively $\bm \varphi$) and with a Lagrangian $\L$. Let us split the Lagrangian into two parts
\begin{equation}
\L=\sqrt{\abs{g}}\left[\L_1(\bm \varphi)+\L_2(\bm\varphi)\right],\label{eq:original lagrangian}
\end{equation}
where $g_{\mu\nu}$ with determinant $g$ is the metric of spacetime. Our results will not depend on the nature of this splitting, but the most useful case will be when $\L_1$ contains the kinetic terms and $\L_2$ contains the potential and interaction terms.

We now add to our theory a vector field $B^\mu$ and consider a new Lagrangian
\begin{equation}
\L'=\sqrt{\abs{g}}\left[\L_1-\hf \frac{M^4}{\L_2}\nabla_\mu B^\mu\nabla_\nu B^\nu\right],\label{eq:lifted lagrangian}
\end{equation}
where $M$ is an arbitrary mass scale, introduced to keep dimensions consistent. Note that we can always set $M=M_{Pl}$ through an appropriate redefinition of $B^\mu$.

Varying~\eqref{eq:lifted lagrangian} with respect to $B^\mu$ gives the equation of motion
\begin{equation}
\partial_\mu\fb{M^2\nabla_\nu B^\nu}{\L_2}=0,
\end{equation}
which implies
\begin{equation}
A=\frac{M^2\nabla_\nu B^\nu}{\L_2}
\end{equation}
is a constant of motion.

Varying~\eqref{eq:lifted lagrangian} with respect to $\varphi^i$ yields the other equation of motion
\begin{equation}
\nabla_\mu\fb{\partial\L_1}{\partial(\partial_\mu\varphi^i)}-\frac{\partial\L_1}{\partial\varphi^i}+\nabla_\mu\left(\frac{A^2}{2}\frac{\partial\L_2}{\partial(\partial_\mu\varphi^i)}\right)-\frac{A^2}{2}\frac{\partial\L_2}{\partial\varphi^i}=0.\label{eq:eom A}
\end{equation}
As we have just seen, $A$ is a constant. If we choose that constant to be $A=\pm\sqrt{2}$ we see that~\eqref{eq:eom A} reduces to
\begin{equation}
\nabla_\mu\fb{\partial(\L_1+\L_2)}{\partial(\partial_\mu\varphi^i)}-\frac{\partial(\L_1+\L_2)}{\partial\varphi^i}=0.
\end{equation}
These are exactly the equations of motion for Lagrangian~\eqref{eq:original lagrangian}. Thus, the theory described by Lagrangian~\eqref{eq:original lagrangian} and the theory described by Lagrangian~\eqref{eq:lifted lagrangian} yield exactly the same classical predictions.

We can use this result to describe any homogeneous scalar field theory in a geometric way, as was shown in~\cite{Finn:2018cfs,Finn:2018krt}. Such a theory will have a Lagrangian of the form
\begin{equation}
\L=\hf k_{ij}(\bm\varphi)\dot{\varphi}^i\dot{\varphi}^j-V(\varphi).\label{eq:scalar lagrangian}
\end{equation}
We now wish to apply the Eisenhart lift to this theory. Since working with homogeneous fields is equivalent to working in one dimension, the field $B^\mu$ has only one component $B^0$, which we can treat as another scalar field. For consistency with previous work we shall relabel this field $B^0\equiv\chi$. 

Taking $\L_1=\hf k_{ij}\dot{\varphi}^i\dot{\varphi}^j$ and $\L_2=-V(\varphi)$ we arrive at the following equivalent Lagrangian:
\begin{equation}
\L'=\hf k_{ij}\dot{\varphi}^i\dot{\varphi}^j+\hf \frac{M^4}{V}\dot{\chi}^2=\hf G_{AB}\dot{\phi}^A\dot{\phi}^B.\label{eq:scalar lifted lagrangian}
\end{equation}
Here $\phi^A\equiv\{\varphi^i,\chi\}$, the indices $A$ and $B$ run from 1 to $N+1$ and
\begin{equation}
G_{AB}\ \equiv\ 
\begin{pmatrix}  
k_{ij}	&	0\\
0		&	\dfrac{M^4}{V}
\end{pmatrix}.\label{eq:scalar metric}
\end{equation}

The Lagrangian~\eqref{eq:scalar lifted lagrangian} describes a system that follows the geodesics of the $N+1$ dimensional field-space manifold with metric~\eqref{eq:scalar metric}. We can therefore describe any theory of the form~\eqref{eq:scalar lagrangian} in a purely geometric manner using this manifold.

\section{Inflation in a Bianchi~I Universe}
We shall study the theory of a single minimally-coupled scalar field in a universe described by Einstein gravity. We therefore take the Lagrangian of the theory to be
\begin{equation}
\L=\sqrt{\abs{g}}\left[-\hf R+\hf(\partial_\mu \varphi)(\partial^\mu \varphi) - V(\varphi)\right],\label{eq:scalar tensor action}
\end{equation}
where $R$ is the Ricci scalar.

As discussed in the Introduction, we shall restrict our attention to spacetimes of Bianchi~I type. The line element in such a spacetime is given by
\begin{equation}
ds^2=dt^2-a_x^2(t)dx^2-a_y^2(t)dy^2-a_z^2(t)dz^2.\label{eq:bianchi metric}
\end{equation}
Furthermore, we shall take the inflaton field to be homogeneous. With these restrictions the Lagrangian~\eqref{eq:scalar tensor action} becomes
\begin{equation}
\L=-a_x\dot{a}_y\dot{a}_z-a_y\dot{a}_x\dot{a}_z-a_z\dot{a}_x\dot{a}_y+\hf a_x a_y a_z\dot{\varphi}^2-a_x a_y a_z V(\varphi).
\end{equation}
This Lagrangian is of the form~\eqref{eq:scalar lagrangian}. We can therefore introduce a new scalar field $\chi$ and use the Eisenhart lift to construct an equivalent Lagrangian:
\begin{equation}
\L'=\hf G_{AB}\dot{\phi}^A\dot{\phi}^B,\label{eq:lifted inflation}
\end{equation}
where $\phi^A=\left\{a_x,a_y,a_z,\varphi,\chi\right\}$ and
\begin{equation}
G_{AB}=\begin{pmatrix}
0&-a_z&-a_y&0&0\\
-a_z&0&-a_x&0&0\\
-a_y&-a_x&0&0&0\\
0&0&0&a_xa_ya_z&0\\
0&0&0&0&\frac{1}{a_xa_ya_zV(\varphi)}
\end{pmatrix}.\label{eq:metric}
\end{equation}
There is a one-to-one correspondence between trajectories of inflation in a Bianchi~I universe and geodesics of the five-dimensional manifold with coordinates $\phi^A$ and metric $G_{AB}$.

\section{A Measure on Initial Conditions}
Following the procedure from F18, we can use the field-space manifold constructed in the previous section to define a measure on the set of initial conditions for inflation. We start by constructing the phase space manifold for the system described by~\eqref{eq:lifted inflation}. This is a ten-dimensional space with coordinates ${\Phi^\alpha=\{\bm\phi,\dot{\bm\phi}\}}$. As discussed in F18, the natural metric for the phase space manifold is the Sasaki metric~\cite{sasakimetric}:
\begin{equation}\label{eq:phase space metric}
\G_{\alpha\beta}=\left(\begin{array}{cc}
G_{AB}+G_{CD}\Gamma^C_{AE}\Gamma^D_{BF}\dot{\phi}^E\dot{\phi^F}&G_{CB}\Gamma^C_{AD}\dot{\phi}^D\\
G_{AC}\Gamma^C_{DB}\dot{\phi}^D&G_{AB}
\end{array}\right),
\end{equation}
where $\Gamma^A_{BC}=\hf G^{AD}\left(G_{BD,C}+G_{DC,B}-G_{BC,D}\right)$ is the Christoffel symbol for the field-space manifold.

The invariant volume element of the phase space manifold,
\begin{equation}
d\Omega=\sqrt{\det\G}\;d^{10}\Phi=\det G\; d^{10}\Phi,\label{eq:phase space measure}
\end{equation}
provides a natural measure on the initial conditions in this model. As shown in F18, the measure~\eqref{eq:phase space measure} is equivalent to the Liouville Measure~\cite{Liouville1838,Gibbons:1986xk,Gibbons:2006pa}.

The Lagrangian~\eqref{eq:lifted inflation} has five symmetries, which leave the equations of motion invariant. These symmetries are: shifts of $\chi$
\begin{equation}
\chi\to\chi+c,\label{eq:shift chi}
\end{equation}
three spatial dilations,
\begin{align}
&a_i\to ca_i, &\dot{a}_i\to c\dot{a}_i,& &\chi\to c\chi,& &\dot{\chi}\to c\dot{\chi},\label{eq:spatial dilation}
\end{align}
for $i\in \{x,y,z\}$ and time dilation,
\begin{align}
&\dot{a}_i\to c\dot{a}_i\;\;\forall i,& &\dot{\chi}\to c\dot{\chi},& &\dot{\varphi}\to c\dot{\varphi}.\label{eq:time dilation}
\end{align}
In \crefrange{eq:shift chi}{eq:time dilation}, $c$ represents an arbitrary constant.

Because of these symmetries there are redundancies in our description of the initial conditions. Any two sets of initial conditions related by one or more of the transformations~\crefrange{eq:shift chi}{eq:time dilation} are physically indistinguishable and, in fact, represent the same Universe. We will therefore integrate out these symmetries to construct a measure of physically distinct initial conditions, as we did in F18.

In order to achieve this we need to change variables to isolate the redundant degrees of freedom. To this end we define the variables
\begin{align}
H_i\equiv\frac{\dot{a}_i}{a_i},& &H_{\chi}\equiv \frac{\dot{\chi}}{a_x a_y a_z},& &\widetilde{\chi}\equiv\frac{\chi}{a_x a_y a_z}.
\end{align}
Now, only $\widetilde{\chi}$ is affected by the transformation~\eqref{eq:shift chi} and only $a_i$ is affected by the transformation~\eqref{eq:spatial dilation}. Thus these symmetries have been isolated.

We further define 
\begin{align}
H_1&\equiv \frac{1}{3}\left(H_x+H_y+H_z\right),\\
H_2&\equiv \frac{1}{6}\left(2H_x-H_y-H_z\right),\\
H_3&\equiv \frac{1}{\sqrt{12}}\left(H_y-H_z\right),
\end{align}
which simplifies the algebra by diagonalising part of the phase space metric~\eqref{eq:phase space metric}. Note that the isotropic case now corresponds to $H_1=H$, $H_2=H_3=0$.

Finally we isolate the symmetry~\eqref{eq:time dilation} by defining
\begin{align}
&H_1=\frac{1}{\sqrt{6}}\rho\cos\alpha\cos\gamma,\;\;H_2=\frac{1}{\sqrt{6}}\rho\cos\alpha\sin\gamma\cos\delta,\nonumber\\
&H_3=\frac{1}{\sqrt{6}}\rho\cos\alpha\sin\gamma\sin\delta,\\
&H_\chi=\rho\sqrt{V}\sin\alpha\sin\beta,\;\;\;\dot{\varphi}=\rho\sin\alpha\cos\beta,\nonumber
\end{align}
so that only $\rho$ is affected by time dilations. Note that the angles defined above cover the ranges
\begin{align}
\begin{aligned}
\alpha&\in \left[-\frac{\pi}{2},\frac{\pi}{2}\right],&\beta&\in \left[0,2\pi\right],\\
\gamma&\in \left[0,\frac{\pi}{2}\right],&\delta&\in \left[0,2\pi\right].
\end{aligned}
\end{align}

Using the above definitions, initial conditions of inflation in this model are fully described by the initial values of the variables
\begin{equation}
\Phi^\alpha=\{a_x,a_y,a_z,\varphi,\widetilde{\chi},\rho,\alpha,\beta,\gamma,\delta\}.\label{eq:coordinates}
\end{equation}
Of these, $\widetilde{\chi}$, $a_x$, $a_y$, $a_z$ and $\rho$ correspond to redundancies of description since their initial values can be arbitrarily changed by the symmetry transformations~\crefrange{eq:shift chi}{eq:time dilation}. We will therefore integrate out these degrees of freedom.

This means that the physically distinct sets of initial conditions can be parametrised by $\varphi$, $\alpha$, $\beta$, $\gamma$ and $\delta$. Of these, $\varphi$, $\alpha$ and $\beta$ were used in F18 and control the initial inflaton field value, the initial expansion rate and the initial inflaton field velocity, respectively. In addition to those we now have $\gamma$, which controls the total degree of initial anisotropy and $\delta$, which controls the direction of that anisotropy.

There is one additional consideration we must take into account. Varying the action~\eqref{eq:scalar tensor action} with respect to $g_{00}$ (also known as the lapse) yields the Hamiltonian constraint, which can be expressed in the variables~\eqref{eq:coordinates} as
\begin{equation}
{\cal H}=\hf a_xa_ya_z\rho^2\left[-\cos^2\alpha\cos(2\gamma)+\sin^2\alpha\right]=0,\label{eq:Hamiltonian constraint}
\end{equation}
where ${\cal H}$ is the Hamiltonian of the theory (which is equal to the Lagrangian for a kinetic-only theory such as~\eqref{eq:lifted inflation}). This is an algebraic constraint on the variables~\eqref{eq:coordinates} and so describes a nine-dimensional submanifold of the phase space, which we call the Hamiltonian hypersurface. Only configurations that lie on the Hamiltonian hypersurface are physically allowed initial conditions for inflation and thus we should use a measure based on this submanifold, not the full phase space.

Following F18 we take the metric on the Hamiltonian hypersurface to be that induced on it by virtue of being embedded in the phase-space manifold. Explicitly, we choose a set of coordinates $\widetilde{\Phi}^a$ (collectively $\widetilde{\bm\Phi}$) on the Hamiltonian hypersurface and encode the embedding through $\Phi^\alpha =F^\alpha(\widetilde{\bm \Phi})$. The induced metric is then given by
\begin{equation}
\widetilde{\G}_{ab}=\frac{\partial F^\alpha}{\partial {\widetilde \Phi}^a}\frac{\partial  F^\beta}{\partial {\widetilde \Phi}^b}\G_{\alpha\beta}.\label{eq:induced metric}
\end{equation}
We therefore take as our measure of initial conditions the invariant volume element on the Hamiltonian hypersurface
\begin{equation}
d\widetilde{\Omega}=\sqrt{\det\widetilde{\G}}d^9\widetilde{\bm\Phi}.
\end{equation}

We choose to use~\eqref{eq:Hamiltonian constraint} to eliminate the variable $\alpha$ and thus describe the Hamiltonian hypersurface using the coordinates
\begin{equation}
\widetilde{\Phi}^a=\{a_x,a_y,a_z,\varphi,\widetilde{\chi},\rho,\beta,\gamma,\delta\}.\label{eq:tilde coordinates}
\end{equation}
The embedding is then described by
\begin{equation}
F^\alpha=\{a_x,a_y,a_z,\varphi,\widetilde{\chi},\rho,\arctan(\sqrt{\cos(2\gamma)}),\beta,\gamma,\delta\}.
\end{equation}
Note, however, that $F^\alpha$ only exists if
\begin{equation}
\gamma\leq\frac{\pi}{4}.
\end{equation}
Therefore, as well as fixing the value of $\alpha$, the Hamiltonian constraint also restricts the allowed degree of anisotropy.

With this in mind we can proceed to calculate the induced metric on the Hamiltonian hypersurface using~\eqref{eq:induced metric}. However, as in F18, we find this metric to be singular with ${\widetilde{\G}_{6a}=\widetilde{\G}_{a6}=0}$ for all values of the index~$a$. Here $6$ refers to the $\rho$ coordinate. We must therefore introduce a regularization technique in order to obtain a sensible measure.

We will use the following, parametrization independent, regularization method which was also used in F18. We consider a submanifold very close to the Hamiltonian hypersurface where ${\cal H}=\epsilon$ and calculate the volume element on this hypersurface before taking the limit $\epsilon\to0$.

Let us denote the induced metric on the surface ${\cal H}=\epsilon$ by $\widetilde{\G}_{ab}(\epsilon)$. Then the invariant volume element on this surface is
\begin{equation}
d\widetilde{\Omega}(\epsilon)=\sqrt{\det\widetilde{\G}(\epsilon)}d^9\widetilde{\bm\Phi}\approx\sqrt{\epsilon}\sqrt{\adj[\widetilde{\G}(0)]^{\alpha\beta}\left.\frac{d\widetilde{\G}_{\alpha\beta}}{d\epsilon}\right|_{\epsilon=0}}d^9\widetilde{\bm\Phi}.\label{eq:epsilon measure}
\end{equation}
Where $\adj[\widetilde{\G}]^{\alpha\beta}$ is the adjugate of $\widetilde{\G}^{\alpha\beta}$ and we have used Jacobi's identity~\cite{Jacobi:1841kw} to evaluate the derivative of the determinant. We see that the volume element is proportional to $\sqrt{\epsilon}$ and is thus singular when $\epsilon\to0$ as expected. However, this overall factor will drop out when the measure is properly normalised and we can safely ignore it. We can therefore take the limit in a sensible fashion at which point the approximation in~\eqref{eq:epsilon measure} becomes exact.

We perform this calculation using the variables~\eqref{eq:tilde coordinates} and find
\begin{equation}
\lim_{\epsilon\to0}\;d\widetilde{\Omega}(\epsilon)=\sqrt{\frac{\epsilon}{2}}a_x^3a_y^3a_z^3\rho^2\frac{\sin(\gamma)}{\cos^3(\gamma)}\frac{1}{\sqrt{V(\varphi)}}d^9\widetilde{\Phi}.
\end{equation}
As explained earlier, the initial values of $\widetilde{\chi}$, $a_x$, $a_y$, $a_z$ and $\rho$ are redundant degrees of freedom and so we integrate them out to obtain a measure on the physically distinguishable initial conditions. We therefore obtain the main result of this Letter:
\begin{equation}
dP=\frac{1}{\N}\frac{\sin(\gamma)}{\cos^3(\gamma)}\frac{1}{\sqrt{V(\varphi)}}d\varphi\,d\beta\,d\gamma\,d\delta,\label{eq:Measure}
\end{equation}
where $dP$ is the measure on the initial conditions and 
\begin{align}
\N&=\int_0^{2\pi}d\beta\int_{0}^{\frac{\pi}{4}}d\gamma\int_0^{2\pi}d\delta\int_{-\infty}^\infty d\varphi\,\left[ \frac{\sin(\gamma)}{\cos^3(\gamma)}\frac{1}{\sqrt{V(\varphi)}}\right]\nonumber\\
&=2\pi^2\int_{-\infty}^\infty \frac{1}{\sqrt{V(\varphi)}}\,d\varphi
\end{align}
is a normalisation constant.

Notice that, just as in F18, the normalisation constant $\N$ is finite provided the inflationary potential diverges quicker than $\varphi^2$ as $\varphi\to\infty$ and there exists a non-zero cosmological constant. Therefore the measure~\eqref{eq:Measure} is well defined and requires no regularisation for this class of inflationary theories.

\section{Conclusions}
We have used the Eisenhart lift to describe inflation in a Bianchi~I universe as the geodesic motion on a five-dimensional field-space manifold with metric~\eqref{eq:metric}. The tangent bundle of that manifold, equipped with a natural metric, provides a phase space manifold that describes all possible sets of initial conditions for inflation in a Bianchi~I universe.

We have shown that, once the Hamiltonian constraint has been taken into account and all redundant degrees of freedom have been integrated out, the total volume of this phase space manifold is finite for the same class of theories found in F18. We can therefore employ Laplace's principle of indifference~\cite{laplace1952philosophical} to argue that the initial conditions should be uniformally distributed over this manifold. This results in the well-defined, normalised measure~\eqref{eq:Measure}.

The measure~\eqref{eq:Measure} factorises into two parts
\begin{equation}
dP=dP_{\rm{FRW}}\;dP_{\rm{anis}},
\end{equation}
where
\begin{equation}
dP_{\rm{FRW}}=\frac{\pi}{\N}\frac{1}{\sqrt{V(\varphi)}}d\varphi\,d\beta\label{eq:FRW measure}
\end{equation}
is the measure for the initial conditions in an FRW universe and
\begin{equation}
dP_{\rm{anis}}=\frac{1}{\pi}\frac{\sin(\gamma)}{\cos^3(\gamma)}d\gamma\,d\delta\label{eq:anis measure}
\end{equation}
is the measure for the anisotropies. The normalisation constants are chosen so that the measures~\eqref{eq:FRW measure} and~\eqref{eq:anis measure} are individually normalised. This separation allows us to analyse independently the initial inflaton field configuration and the initial spacetime geometry.

The measure~\eqref{eq:FRW measure} is identical to the one found in F18 and thus many of the same conclusions will still hold. In particular, provided $\N$ is finite, we find that the region of phase space in which the inflaton field is displaced from its minimum takes up a significantly smaller fraction of the measure than the region of phase space with the inflaton at or near its minimum. Thus, under this measure, initial conditions that lead to significant inflation are indeed finely tuned. A full analysis of this measure, including quantitative results, can be found in F18.

\begin{figure}
\includegraphics[width=0.5\textwidth]{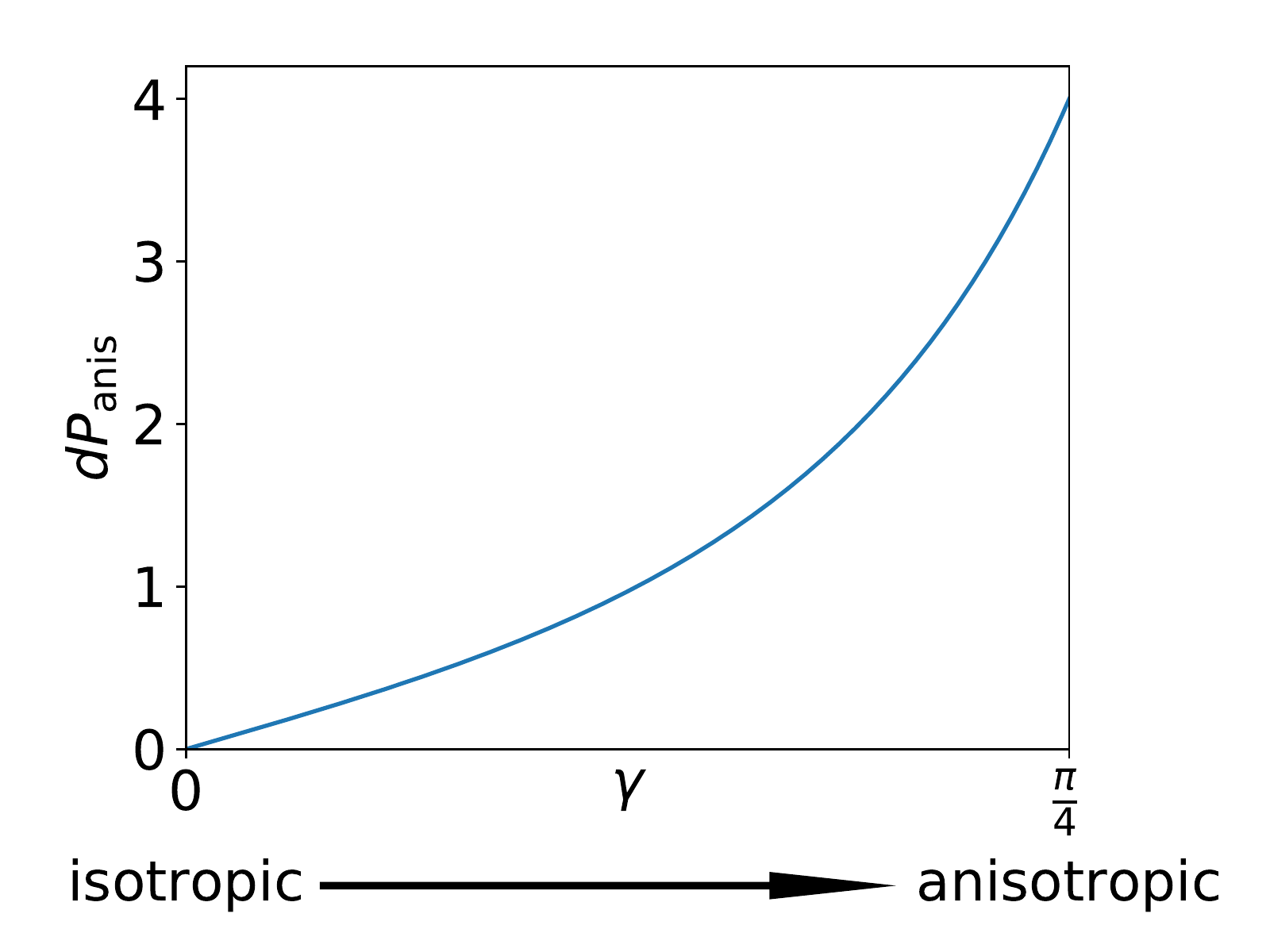}
\caption{Measure on the initial anisotropy of the Universe as given by~\eqref{eq:anis measure}.\label{fig:prob_plot}}
\end{figure}

The measure on initial spacetime geometry, which in our case is restricted to an initial anisotropy parametrised by~$\gamma$ and~$\delta$, is given by~\eqref{eq:anis measure}. We see that this measure is independent of the inflationary potential and is uniform over $\delta$. The measure can therefore be considered as a distribution on the degree of anisotropy present in the initial state of the Universe, which is independent of the inflationary model. This distribution is shown in Figure~\ref{fig:prob_plot}.

As we can see in Figure~\ref{fig:prob_plot}, the measure slightly favours anisotropic universes over isotropic ones. However, if inflation does occur, it will dilute any initial anisotropy by an exponential amount. Therefore anisotropy will only be observable today if it was initially exponentially large. Such initial conditions represent a tiny fraction of the measure~\eqref{eq:anis measure}, despite the mild enhancement of anisotropic universes. We therefore see that the inclusion of anisotropy has a negligible impact on the results of F18.

Relaxing the assumption of isotropy does not solve the fine tuning issues observed in F18. In the manifold of all possible initial conditions for a single scalar field in a Bianchi~I universe, the set that allows $N>60$ e-foldings of inflation represents only a tiny fraction. It is therefore far from clear that inflation truly solves the fine-tuning puzzles that it was designed for.
\\[12pt]
The author would like to thank Sotirios Karamitsos, Apostolos Pilaftsis and David Sloan for useful comments and discussion. KF is supported by the University of Manchester through the President's Doctoral Scholar Award.

\end{document}